\documentclass[aps,prl,twocolumn,superscriptaddress]{revtex4}
\usepackage{graphicx}
\usepackage{color}
\usepackage[tight]{units}

\definecolor{light-gray}{rgb}{0.8,0.8,0.8}

\begin{document}

\title{Electron-phonon coupling in the spin-split valence band of single layer WS$_2$}
\author{Nicki Frank Hinsche}
\affiliation{Center for Atomic-scale Materials Design, Department of Physics, Technical University of Denmark, DK-2800 Kgs. Lyngby, Denmark}
\email{nickih@fysik.dtu.dk}
\author{Arlette S. Ngankeu}
\author{Kevin Guilloy}
\author{Sanjoy K. Mahatha}
\author{Antonija Grubi\v{s}i\'{c} \v{C}abo}
\author{Marco Bianchi}
\author{Maciej Dendzik}
\author{Charlotte E. Sanders}
\author{Jill A. Miwa}
\affiliation{Department of Physics and Astronomy, Interdisciplinary Nanoscience Center (iNANO), Aarhus University, 8000 Aarhus C, Denmark}
\author{Harsh Bana}
\affiliation{Physics Department, University of Trieste, Via Valerio 2, 34127 Trieste, Italy}
\author{Elisabetta Travaglia}
\affiliation{Physics Department, University of Trieste, Via Valerio 2, 34127 Trieste, Italy}
\author{Paolo Lacovig}
\affiliation{Elettra - Sincrotrone Trieste S.C.p.A., AREA Science Park, S.S. 14 km 163.5, 34149 Trieste, Italy}
\author{Luca Bignardi}
\affiliation{Elettra - Sincrotrone Trieste S.C.p.A., AREA Science Park, S.S. 14 km 163.5, 34149 Trieste, Italy}
\author{Rosanna Larciprete}
\affiliation{CNR-Institute for Complex Systems, Via Fosso del Cavaliere 100, 00133 Roma, Italy}
\author{Alessandro Baraldi}
\affiliation{Physics Department, University of Trieste, Via Valerio 2, 34127 Trieste, Italy}
\affiliation{Elettra - Sincrotrone Trieste S.C.p.A., AREA Science Park, S.S. 14 km 163.5, 34149 Trieste, Italy}
\affiliation{IOM-CNR, Laboratorio TASC, AREA Science Park, S.S. 14 km 163.5, 34149 Trieste, Italy}
\author{Silvano Lizzit}
\affiliation{Elettra - Sincrotrone Trieste S.C.p.A., AREA Science Park, S.S. 14 km 163.5, 34149 Trieste, Italy}
\author{Kristian Sommer Thygesen}
\affiliation{Center for Atomic-scale Materials Design, Department of Physics, Technical University of Denmark, DK-2800 Kgs. Lyngby, Denmark}
\author{Philip Hofmann}
\email{philip@phys.au.dk}
\affiliation{Department of Physics and Astronomy, Interdisciplinary Nanoscience Center (iNANO), Aarhus University, 8000 Aarhus C, Denmark}

\date{\today}
\begin{abstract}
The absence of inversion symmetry leads to a strong spin-orbit splitting of the upper valence band of semiconducting single layer transition metal dichalchogenides  such as MoS$_2$ or WS$_2$. This permits a direct comparison of the electron-phonon coupling strength in states that only differ by their spin. Here, the electron-phonon coupling in the valence band maximum of single-layer WS$_2$ is studied by first principles calculations and angle-resolved photoemission. The coupling strength is found to be drastically different for the two spin-split branches, with calculated values of $\lambda_K=$0.0021 and 0.40 for the upper and lower spin-split valence band of the free-standing layer, respectively. This difference is somewhat reduced when including scattering processes involving the Au(111) substrate present in the experiment and the experimental results confirm the strongly branch-dependent coupling strength.
\end{abstract}
\maketitle

\begin{figure}
\includegraphics[width=0.45\textwidth]{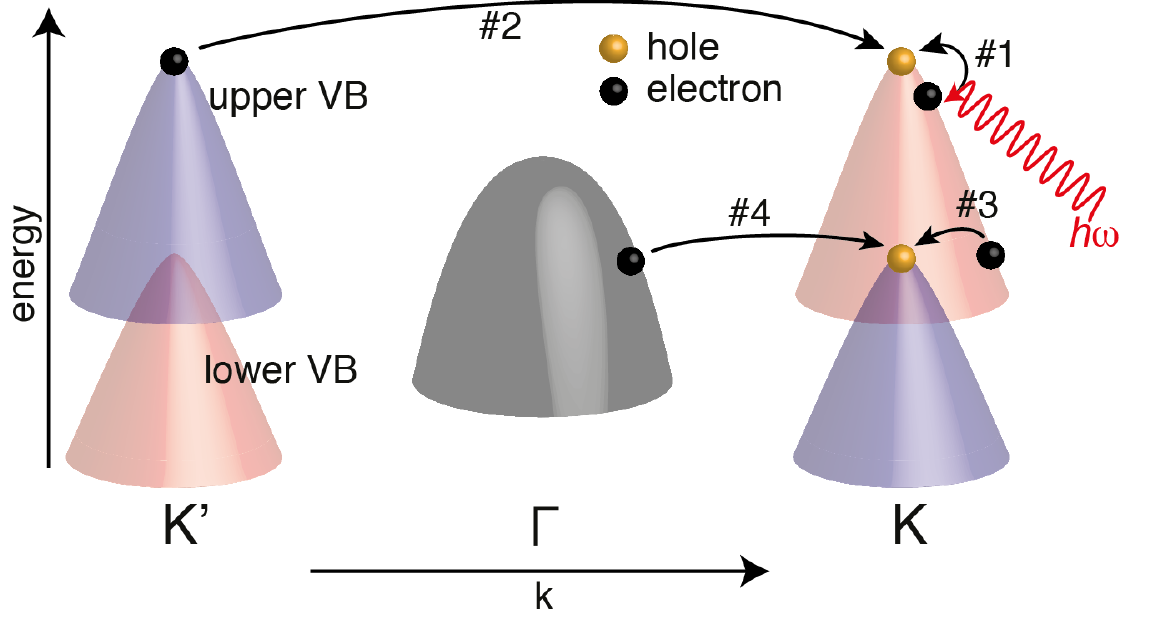}\\
\caption{(Color online) Schematic valence band for SL WS$_2$ with the absolute valence band (VB) maxima at K and K$^{\prime}$ and a local maximum at $\Gamma$. The VB at K and K$^{\prime}$ are spin-split with the color indicating the spin direction. The VB at $\Gamma$ is spin-degenerate. El-ph scattering processes to fill holes at the top of the spin-split band at K are illustrated. In process \#1, an electron is scattered into the hole by absorbing a phonon (red line). For the other scattering processes, the phonons are omitted for clarity. }
  \label{fig:1}
\end{figure}

The electronic structure of semiconducting single layer (SL)  transition metal dichalcogenides  with a trigonal prismatic structure (MoS$_2$, WS$_2$) resembles that of graphene in certain respects, but with a sizable band gap \cite{Bollinger:2001aa,Mak:2010aa,Splendiani:2010aa}. On closer inspection, however, an important difference from graphene is the lack of inversion symmetry which, combined with the presence of heavy atoms in the materials, leads to a lifting of the spin-degeneracy in the band structure. This is especially important near the valence band (VB) maximum at K, where the spin splitting can be substantial (see Fig. \ref{fig:1}). This spin-splitting, combined with the two valleys at K and K$^{\prime}$, gives rise to new spin-like quantum degrees of freedom \cite{Xu:2014ac} that could be exploited for storing or transmitting quantum information. Moreover, in hole-doped materials, the valence band's spin texture can have consequences ranging from increased hole mobility to topological superconductivity \cite{Hsu:2017aa}. 

The strong spin-splitting in the band structure also provides a unique opportunity to study the electron-phonon (el-ph) coupling in a system of states that only differ by their spin \cite{Xiao:2012ab}. In this Letter, we report theoretical and experimental results for the el-ph coupling in SL WS$_2$ and find that the coupling strength is strongly dependent on the branch of the spin-split band structure. While this might appear surprising at first glance, it can be explained by a combination of spin-protected scattering and, more importantly, phase space restrictions. Our findings are directly relevant to hole transport in single layers devices and to the yet-to-be-demonstrated superconductivity in hole-doped materials \cite{Costanzo:2016aa}.

The expected el-ph scattering contributions for the VB maximum of SL WS$_2$ are shown in Fig. \ref{fig:1} in a schematic representation of the band structure. The important features for the band structure are the two spin-split VB maxima at K and K$^{\prime}$. Due to time-reversal symmetry, the spin structure is reversed at these two points. In the following, the spin split bands at K will be referred to as the ``upper'' and ``lower'' bands, the ``upper'' band being the absolute VB maximum. Situated energetically between the maxima of the upper and lower bands, there is also a local VB maximum at $\Gamma$ which is spin-degenerate. Consider now how a hole can be filled by an electron with the help of crystal momentum (and energy) of a phonon. A hole near the top of the upper band at K can only be filled by an electron at higher binding energy, a scenario which requires the absorption of a thermally excited phonon. This electron has to come either from the same band at higher binding energy (process \#1) or from the band at K$^{\prime}$ (process \#2). The latter intervalley-scattering process would require a spin flip, or at least a considerable change in the spin-expectation value, making the process extremely unlikely \cite{Fabian:1999cm}. Note that no el-ph scattering processes from the lower band or the band at $\Gamma$ are possible because the energy difference between these bands and the upper band ($\approx \unit[0.43]{eV}$ and $\approx \unit[0.22]{eV}$, respectively) far exceed the maximum phonon energy of $\approx \unit[55]{meV}$ \cite{Molina-Sanchez:2011aa}. The situation is drastically different in the lower band, for which several additional scattering processes are possible---e.g., from the VB at $\Gamma$ (process \#4); or, from a state in the upper branch away from K (process \#3), for which the spin polarization is lower than it is exactly at K \cite{Xiao:2012ab,SMAT}; or from the lower band near K$^{\prime}$ (not shown in the figure). Moreover, all of these processes can also proceed via phonon emission at low temperatures. Based on this simple picture, one would expect a substantially weaker el-ph coupling in the upper band than in the lower band, and this will also be confirmed by our findings. Note that this result is only partly a direct consequence of the state's spin texture, even though the upper and lower bands only differ by the spin part of their wave function. An important consideration is also the restriction of scattering phase  space---i.e., the number of available final states---which can be a dominant factor for the el-ph coupling strength \cite{Gayone:2003aa}.

The electronic structure of SL WS$_2$ and of the Au(111) substrate used in the experiments was calculated from first principles within density functional theory, as implemented in the \textsc{Quantum\-Espresso} code \cite{Giannozzi:2009hx}. The phonon properties were obtained within density functional perturbation theory using the same package. Relativistic effects, e.g. spin-orbit coupling, were treated self-consistently and are accounted for in all calculations. The unreconstructed Au(111) substrate was modelled by a 9-layer slab. On the basis of the fully relaxed atomic positions, the electronic spectral function was calculated and the relative band alignments of the SL WS$_2$ and the Au(111) substrate were determined \cite{theorydetails,SMAT}. 

Based on the electronic and vibrational properties, the el-ph coupling was obtained within a modified version of the \textsc{EPW} code \cite{Ponce:2016dv}, either for the free-standing SL WS$_2$ or for the states in the SL in the presence of the Au(111) substrate. An improved tetrahedron Fermi-surface-adaptive integration scheme based on the Wannier-interpolated electron-phonon matrix elements was applied \cite{Zahn:2011ci,Kawamura:2014cr,Assmann:2016wi,Rittweger:2017}.

The \textit{a priori} state-dependent, temperature-dependent phonon-induced electronic linewidth $\Gamma_{n\mathbf{k}}$ is closely related to the imaginary part of the lowest order el-ph self-energy $\Sigma_{n\mathbf{k}}''$ for the Bloch state of energy 
$\varepsilon_{n\mathbf{k}}$ at band $n$ and momentum $\mathbf{k}$ and given by 
\begin{eqnarray}\label{equ:ellw}
\Gamma_{n\mathbf{k}}(T) =  2\Sigma_{n\mathbf{k}}''(T) = 2\pi \sum_{m\nu} \int_{\rm BZ} \frac{d\mathbf{q}}{\Omega_{\rm BZ}} | g_{mn,\nu}(\mathbf{k,q}) |^2 \nonumber \\
    \times   \Big\{\big[n_{\mathbf{q}\nu}(T)+f_{m\mathbf{k+q}}(T)\big]\delta(\varepsilon_{n\mathbf{k}}-\varepsilon_{m\mathbf{k+q}} +\omega_{\mathbf{q}\nu}) \nonumber \\
+ \big[n_{\mathbf{q}\nu}(T)+1-f_{m\mathbf{k+q}}(T)\big]\delta(\varepsilon_{n\mathbf{k}} -\varepsilon_{m\mathbf{k+q}} - \omega_{\mathbf{q}\nu})\Big\},
\end{eqnarray}
where $g_{mn,\nu}(\mathbf{k,q})$ is the el-ph scattering matrix element; $n$ and $f$ are Bose and Fermi functions, respectively; $\varepsilon$ and $\omega$ are non-interacting electron and phonon energies, respectively. To ensure convergence up to $\unit[1.5 \times 10^6]{}$ random final states and their corresponding Fourier-interpolated $\omega_{\mathbf{q}\nu}$ and Wannier-interpolated $g_{mn,\nu}(\mathbf{k,q})$ were taken into account. 
The electron-phonon coupling strength $ \lambda_{n\mathbf{k}}$ is essentially counting the possible scattering processes for a chosen initial state $(\mathbf{k},n)$ into possible final states, weighted by a squared matrix element:
\begin{eqnarray}\label{equ:lambda}
   \lambda_{n\mathbf{k}} = \sum_{m\nu} \int_{\rm BZ} \frac{d\mathbf{q}}{\Omega_{\rm BZ} \omega_{\mathbf{q}\nu}} | g_{mn,\nu}(\mathbf{k,q}) |^2 \nonumber \\
    \times  \delta(\varepsilon_{n\mathbf{k}}-\varepsilon_{m\mathbf{k+q}} \pm \omega_{\mathbf{q}\nu}).
\end{eqnarray} 
Note that $ \lambda_{n\mathbf{k}}$ has the character of an energy- and $k$-dependent general coupling constant and is thus a more general quantity than the mass-enhancement parameter at the Fermi surface of a metal \cite{Grimvall:1981aa}. 
A signature of the electron-phonon interaction is the linear temperature dependence of 
$\Gamma_{n\mathbf{k}}$ at high temperatures. The latter is a direct consequence of the T dependence of the phonon occupation numbers---i.e., $n_{\mathbf{q}\nu}(T \rightarrow \infty) \propto \nicefrac{k_{B}T}{\omega_{\mathbf{q}\nu}}$. Comparing equ.~\ref{equ:ellw} and \ref{equ:lambda} one readily obtains $\Gamma_{n\mathbf{k}}(T)=2\pi \lambda_{n\mathbf{k}} k_B T$ \cite{Eiguren:2003ed}.

In the experimental part of the current study, SL WS$_2$ was epitaxially grown on Au(111) following a well-established procedure of W evaporation onto the clean Au surface in a background pressure of H$_2$S \cite{Lauritsen:2007aa,Miwa:2015aa,Dendzik:2015aa}. A careful optimization of the growth conditions, guided by the minimization of the core level spectra linewidth, resulted in higher crystal quality than in Ref. \cite{Dendzik:2015aa}, as also reflected in the narrower VB features in the present study. Samples were synthesized at the SuperESCA beamline at ELETTRA \cite{Baraldi:2003ab} and ARPES data were collected on the SGM-3 beamline of ASTRID2 \cite{Hoffmann:2004aa}.  The energy and angular resolution were better than 30~meV and 0.2$^{\circ}$, respectively.

\begin{figure}
\includegraphics[width=0.5\textwidth]{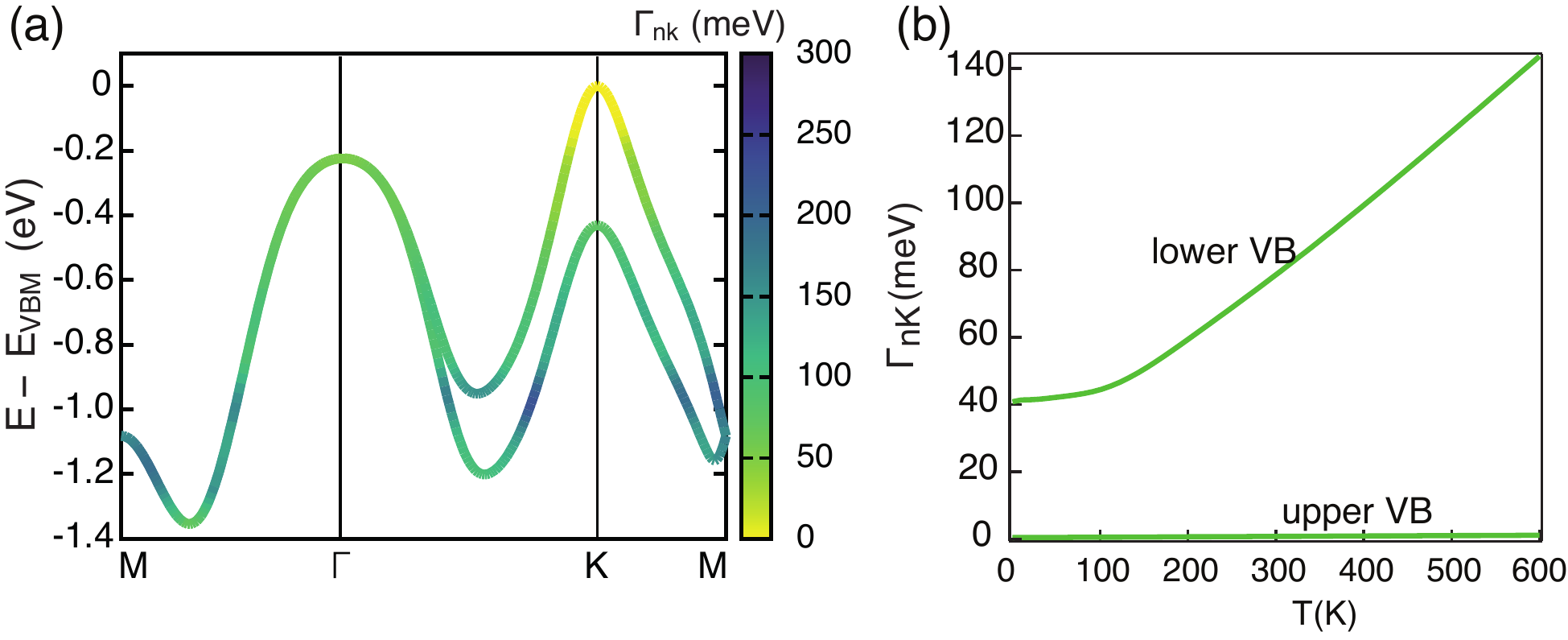}\\
\caption{(Color online) (a) Calculated valence band structure for free-standing SL WS$_2$. The color coding of the bands represents the el-ph coupling induced linewidth of the state at 300~K.  (b) Calculated temperature-dependent linewidth for the spin-split branches of the valence band at K.  }
  \label{fig:2}
\end{figure}

The calculations  for free-standing SL WS$_2$ are presented in Fig. \ref{fig:2}. Fig. \ref{fig:2}(a) shows the calculated VB dispersion with a color coding that represents the linewidth at $T=$300~K, calculated using equ. \ref{equ:ellw}. This figure already confirms the  simple picture discussed in connection with Fig. \ref{fig:1}: We find a very narrow linewidth for the top of the upper band near K (a long hole lifetime) and a substantially increased linewidth for the top of the lower band. Phase space considerations also appear to be significant for other states at higher binding energies, with particularly short lifetimes close to critical points in the density of states. 

Fig. \ref{fig:2}(b) shows the calculated linewidth for the upper and lower bands at K as a function of temperature, demonstrating once again the much stronger el-ph coupling for the lower band. The linear high-temperature regime appears to be reached above $\approx 300$~K, and a linear fit to the linewidth in the experimentally accessible temperature range between 300~K and 600~K yields $\lambda_K=\unit[0.0021]{}$ ($\lambda_K=\unit[0.40]{}$) for upper (lower) VB, respectively. It turns out, however, that the dependence is not completely linear, due to high energy phonons not being equally occupied in this temperature range, and the corresponding values from a fit for even higher temperatures ($\unit[550-750]{K}$) are slightly higher, with $\lambda=\unit[0.0025]{}$ ($\lambda=\unit[0.42]{}$).

\begin{figure}
\includegraphics[width=0.5\textwidth]{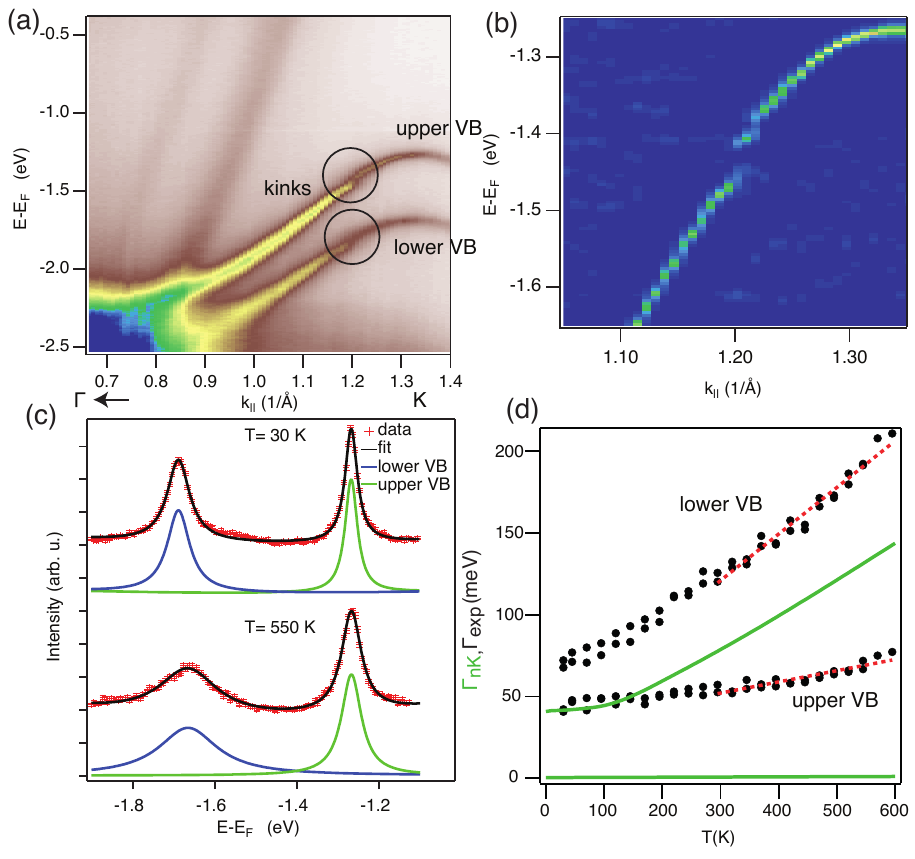}\\
\caption{(Color online) (a) Photoemission intensity along $\Gamma$-K at 30~K. The positions of kinks in the WS$_2$ bands are marked. (b) Curvature of the photoemission data presented in (a) \cite{Zhang:2011aa}, showing the kink in the upper valence band. (c) EDCs through K at low and high temperature. The markers are the data points. The resulting fit with a polynomial background and two Lorentzian peaks are also shown. (d) Temperature-dependent Lorentzian linewidth of the two bands in an energy distribution curve through K (markers). The solid lines show the calculation for the free-standing layer from Fig. \ref{fig:2}(b). The dashed lines are a fit of the experimental data at high temperature ($T>300$~K.)}
  \label{fig:3}
\end{figure}

Figure \ref{fig:3}(a) shows the experimentally determined band structure for high-quality SL WS$_2$ on Au(111) close to the K point, with the two well-separated spin-split branches of the VB. Since this part of the band structure is found in a projected bulk band gap of Au(111) \cite{Takeuchi:1991aa}, there is little direct hybridization between the SL WS$_2$ and substrate bands, and the states are very narrow. The high quality of the sample reveals some hitherto undetected details. In particular, a kink-like deviation from the nearly parabolic dispersion is visible at energies of 174$\pm$14~meV and 124$\pm$18~meV below the top of the upper and lower VB branches, respectively (for a magnification see Fig. \ref{fig:3}(b)). Such kinks are often indicative of strong el-ph coupling, and while they normally occur close to the Fermi energy \cite{Lanzara:2001aa}, they can also be found at higher binding energy such as near the top of the $\sigma$-band of graphene \cite{Mazzola:2013aa,Mazzola:2017aa}. Here, el-ph coupling can be ruled out as the cause of the kink, as it occurs far outside the phonon energy window. Instead, we assign the kink to a minigap opening at the new Brillouin zone boundary caused by the moir\'e superstructure formed between SL WS$_2$ and Au(111). Note that the minigaps  do not affect the band structure exactly at K \cite{SMAT}. 

The effect of el-ph coupling on the two spin-split branches can be determined by analyzing energy distribution curves (EDCs) through the K point for data taken at different temperatures.  Fig. \ref{fig:3}(c) shows a comparison of data taken at 30~K and 550~K \cite{SMAT}. Even without any detailed analysis, it is clear that the temperature-induced broadening of the lower branch is substantially stronger than for the upper branch, in qualitative agreement with the theoretical results. For a more quantitative analysis, EDCs taken over a wide temperature range are fitted using a polynomial background and two Lorentzian peaks \cite{SMAT}, and the resulting linewidth  $\Gamma (T)$ is plotted in Fig. \ref{fig:3}(d) together with the theoretical result from Fig. \ref{fig:2}(b) \cite{Hofmann:2009ab}. The experimental linewidth is not expected to be identical to the theoretically calculated value of  equ. \ref{equ:ellw} because it contains contributions of electron-electron and electron-defect scattering. These can be significant but they are generally independent of the temperature. When comparing calculation and experiment, one should hence allow for a temperature-independent offset. 

From the data in Fig. \ref{fig:3}(d), the coupling constant $\lambda$ can be estimated in the same way as previously used for the theoretical data: A linear fit of the data for high temperatures ($T>300$~K, dashed lines in the figure) yields coupling constants of $\lambda_K \approx$ 0.52 and 0.13 for the lower and upper  VB, respectively. Note, however, the experimental linewidth is not reaching a truly linear regime and therefore the resulting $\lambda$ depends on the temperature range chosen for the fit. Therefore, rather than attempting an accurate determination of $\lambda$ from the experimental data, we concentrate on a direct comparison of measured and calculated linewidths. 
 
For the lower band, a good agreement between the experimental and calculated linewidths is found, especially when allowing for a temperature-independent offset between the curves. Indeed, the calculated slope at high temperature can  be seen to be very similar to the experimental result. For the upper band, the agreement is less satisfying. Even when adding a constant offset to the calculated linewidth, the  experimental data still shows a noticeable slope (reflected by $\lambda \approx 0.13$), in contrast to the nearly flat calculated curve.

An aspect that has so far not been considered is a possible involvement of the Au(111) substrate states. The presence of the substrate enables new decay channels via el-ph, electron-electron and electron-defect scattering. Scattering into the substrate is known to be a significant process for the decay of excited carriers in the conduction band \cite{Antonija-Grubisic-Cabo:2015aa}.  Including the substrate states in the calculation of $\Sigma''$ and $\lambda$ is straight-forward because it merely requires that the sums in equs. \ref{equ:ellw} and \ref{equ:lambda} be extended over the Au(111) states \cite{SMAT}.  Note, however, that this approach neglects effects of band hybridization between SL WS$_2$ and the Au substrate, as well as the impact of the Au(111) surface state, at the expense of computational feasibility. None of the latter effects is expected to have a major impact on the final results at K, as band hybridization effects were found to mainly occur in the vicinity of $\Gamma$ at high binding energies. However, the hybridization in the possible final states for scattering processes might still affect the quantitative results. The Au(111) surface state is found to be energetically well above the VB maximum and should thus not play a role here \cite{Dendzik:2015aa}.

\begin{figure}
\includegraphics[width=0.5\textwidth]{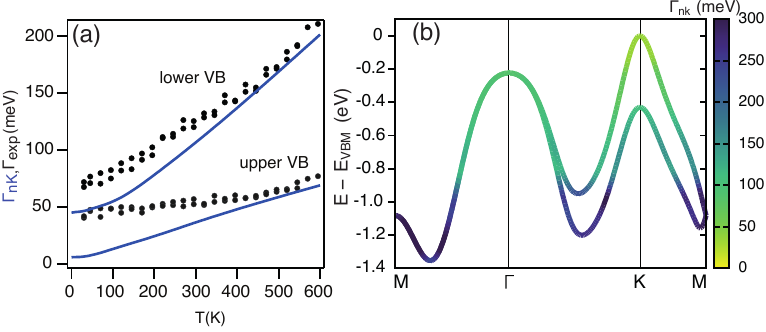}\\
\caption{(Color online) Temperature-dependent linewidth of the spin-split valence band branches at K, showing the experimental data together with a calculation including scattering contributions from the Au(111) substrate. (b) Band structure with encoded linewidths, as in Fig. \ref{fig:2}(b), but including the effects of the Au(111) substrate. }
  \label{fig:4}
\end{figure}

When including the substrate's contribution to the el-ph coupling, a linear fit of the calculated linewidth between 300 and 600~K results in $\lambda_{K}=\unit[0.20]{}$ ($\lambda_{K}=\unit[0.58]{}$) for upper  (lower) VB, respectively. Thus an improved agreement for temperature dependence of the upper band is found, but this happens at the expense of the agreement for the lower band, where the theoretical coupling strength now exceeds the experimental value (see. Fig. \ref{fig:4}(a)).  Fig. \ref{fig:4}(b) shows the VB band structure with a colour-coding of the linewidth, in the same way as in Fig. \ref{fig:2}(b) but involving the Au substrate. Comparing Figs. \ref{fig:2}(b) and \ref{fig:4}(b) reveals that the Au substrate leads to an overall broadening of the states, especially at high binding energies. However, the  energy-dependent changes of the two branches are still very similar, as seen in a detailed plot of their linewidth as a function of binding energy in Ref. \cite{SMAT}. 

While the strong branch dependence of the el-ph coupling is present in both calculations and in the experimental data, and while there is good agreement for the high-temperature slopes of the temperature-dependent linewidths, our model does not provide an accurate quantitative description of the experimental data. This is evident in the entire temperature range shown in Fig. \ref{fig:4}(a). Clearly, experimental results and calculations cannot be reconciled by a simple rigid shift to account for temperature-independent scattering mechanisms. The most likely explanation lies in the simplicity of the model, which requires a relatively small unit cell to be tractable and does not describe the hybridization of SL WS$_2$ and Au(111) adequately.

In summary, our results reveal drastically different el-ph coupling strengths in the spin-split VB states at K in SL WS$_2$. Indeed, holes in the upper band can be expected to have a very long lifetime for a free-standing layer, in contrast to holes in the lower band that are fairly strongly affected by el-ph coupling. The strong spin-orbit interaction in WS$_2$ could easily permit inducing a hole population in the upper band at K without a significant number of carriers in the lower band or in the branch at $\Gamma$. Such a scenario should not only yield high hole mobilities, but also constitute a realization of the system which has been discussed as a candidate for topological superconductivity in Ref. \cite{Hsu:2017aa}. Conventional el-ph mediated superconductivity, on the other hand, would be extremely unlikely in this situation. 

This work was supported by the Danish Council for Independent Research, Natural Sciences under the Sapere Aude program (Grant No. DFF-4002-00029) and by VILLUM FONDEN via the Centre of Excellence for Dirac Materials (Grant No. 11744). NFH received funding within the H.C. \O rsted Programme from the European Union's Seventh Framework Programme and Horizon 2020 Research and Innovation Programme under Marie Sklodowska-Curie Actions grant no. 609405 (FP7) and 713683 (H2020). The Center for Nanostructured Graphene (CNG) is sponsored by the Danish National Research Foundation, Project No. DNRF103.

\end{document}